\documentclass[journal=nalefd,manuscript=letter]{achemso}
\usepackage[version=3]{mhchem}
\usepackage{amsmath}
\usepackage{amsfonts}
\usepackage{amssymb}
\usepackage{graphicx}
\usepackage{color}

\title[] {$p$-GaAs nanowire MESFETs with near-thermal limit gating}

\author{A.R.~Ullah}
\affiliation{School of Physics, University of New South Wales, Sydney NSW 2052, Australia}

\author{F. Meyer}
\affiliation{School of Physics, University of New South Wales, Sydney NSW 2052, Australia}

\author{J.G. Gluschke}
\affiliation{School of Physics, University of New South Wales, Sydney NSW 2052, Australia}

\author{S. Naureen}
\affiliation{Department of Electronic Materials Engineering, Research School of Physics and Engineering, The Australian National University, Canberra ACT 2601, Australia}
\alsoaffiliation[Now at: ]{IRnova AB, Electrum 236, Kista SE-164 40, Sweden}

\author{P. Caroff}
\affiliation{Department of Electronic Materials Engineering, Research School of Physics and Engineering, The Australian National University, Canberra ACT 2601, Australia}
\alsoaffiliation[Now at: ]{Microsoft Station Q, Delft University of Technology, 2600 GA Delft, The Netherlands}

\author{P.~Krogstrup}
\affiliation{Center for Quantum Devices, Niels Bohr Institute, University of Copenhagen, DK-2100 Copenhagen, Denmark}

\author{J.~Nyg{\aa}rd}
\affiliation{Center for Quantum Devices, Niels Bohr Institute, University of Copenhagen, DK-2100 Copenhagen, Denmark}

\author{A.P.~Micolich}
\email{adam.micolich@nanoelectronics.physics.unsw.edu.au}
\affiliation{School of Physics, University of New South Wales, Sydney NSW 2052, Australia}

\date{\today}

\begin{document}

\begin{abstract}
Difficulties in obtaining high-performance $p$-type transistors and gate insulator charge-trapping effects present two major challenges for III-V complementary metal-oxide semiconductor (CMOS) electronics. We report a $p$-GaAs nanowire metal-semiconductor field-effect transistor (MESFET) that eliminates the need for a gate insulator by exploiting the Schottky barrier at the metal-GaAs interface. Our device beats the best-performing $p$-GaSb nanowire metal-oxide-semiconductor field effect transistor (MOSFET), giving a typical sub-threshold swing of $62$~mV/dec, within $4\%$ of the thermal limit, on-off ratio $\sim 10^{5}$, on-resistance $\sim 700$~k$\Omega$, contact resistance $\sim 30$~k$\Omega$, peak transconductance $1.2~\mu$S$/\mu$m and high-fidelity ac operation at frequencies up to $10$~kHz. The device consists of a GaAs nanowire with an undoped core and heavily Be-doped shell. We carefully etch back the nanowire at the gate locations to obtain Schottky-barrier insulated gates whilst leaving the doped shell intact at the contacts to obtain low contact resistance. Our device opens a path to all-GaAs nanowire MESFET complementary circuits with simplified fabrication and improved performance.

\end{abstract}

\maketitle

Modern integrated circuits are heavily reliant on complementary circuit architectures featuring both $p$-type and $n$-type transistors to minimise power consumption~\cite{ChandrakasanJSSC92}. Continued miniaturisation spurred the development of nanowire CMOS,~\cite{ITRS15} at first using carbon nanotubes and Si nanowires~\cite{LuNM07, BarraudIEDM17, MertensIEDM17}, and more recently focussing on III-V nanowires~\cite{WernerssonProcIEEE10, delAlamoNat11, RielMRSBull14, delAlamoJEDS16} integrated on Si towards achieving high performance at low cost. Progress for $n$-type III-V nanowire transistors has fared better than for $p$-type. Near-thermal limit gating has been obtained for $n$-InP~\cite{StormNL11}, $n$-InGaAs~\cite{TomiokaNat12} and $n$-AlGaAs/GaAs~\cite{MorkotterNL15} nanowires, with integration on Si substrates~\cite{TomiokaNat12, SchmidAPL15} and GHz operation~\cite{JohanssonEDL14} also demonstrated. Several significant challenges have impeded progress on the $p$-type counterparts including lower intrinsic carrier mobility and difficulties in the growth, doping and fabrication of high quality gates and ohmic contacts~\cite{WernerssonProcIEEE10,delAlamoJEDS16}.

Candidate materials for $p$-type III-V nanowire transistors include GaSb, GaAs, InAs, InGaAs, InP and InSb. The In-based materials are hard to deploy and often ambipolar because the $p$-type doping needs to compete against sub-surface electron accumulation arising from surface-state pinning of the surface Fermi energy at the conduction band edge and/or a relatively small band-gap~\cite{SorensonAPL08, StormNL11, NilssonAPL10}. GaSb is more favourable for $p$-type nanowire transistors because it is intrinsically $p$-type due to native antisite defects~\cite{LingAPL04, VirkkalaPRB12}. Dey {\it et al.}~\cite{DeyNL12} reported a single InAs/GaSb nanowire CMOS inverter circuit featuring a horizontally-oriented GaSb $p$-MOSFET with sub-threshold swing $S = 400$~mV/dec, peak transconductance $g_m = 3.4~\mu$S$/\mu$m, on-off ratio $\sim10^{1.8}$, on-resistance $R_{on}~>~1.2$~M$\Omega$ and operation frequency up to $10$~kHz without significant fidelity loss. Babadi {\it et al.}~\cite{BabadiAPL17} obtained improved $R_{on}\sim26$~k$\Omega$ from a GaSb nanowire with moderate Zn doping albeit with some loss in sub-threshold swing ($820$~mV/dec). Vertical $p$-GaSb MOSFET arrays have since been developed towards scale-up and applications;~\cite{SvenssonNL15,JonssonEDL18} we will return to these later. A complexity with using GaSb nanowires is the need to grow on GaAs or InAs stems to achieve high quality and yield.~\cite{BorgNano13}

\begin{figure}
\includegraphics[width=12cm]{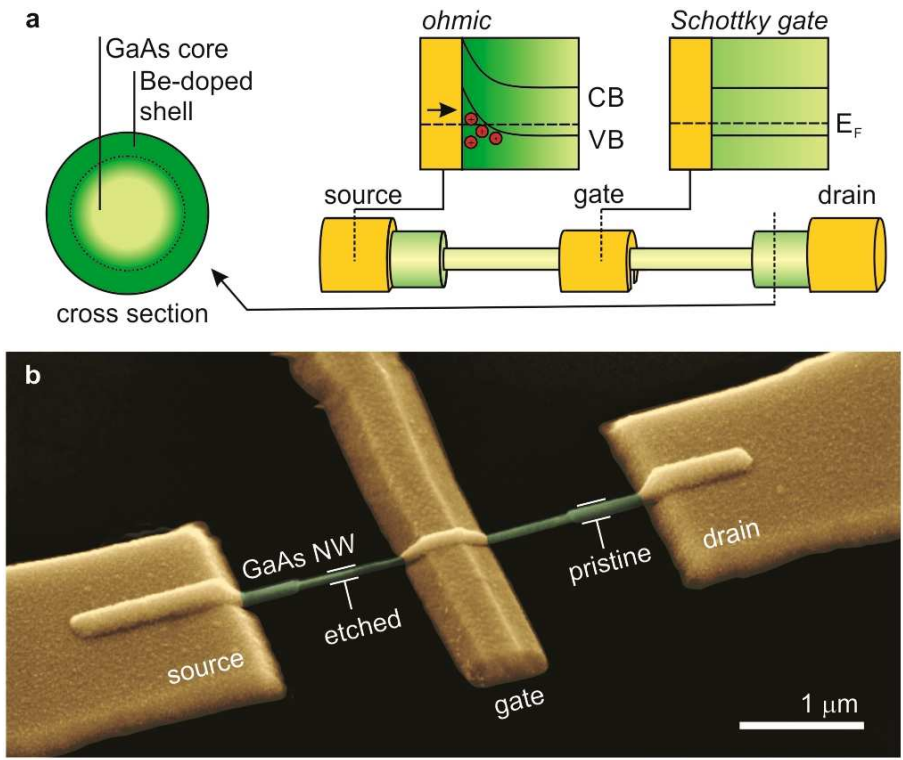}
\vspace{2mm}
\caption{{\bf Device Structure.} {\bf a} Schematic diagram and {\bf b} Scanning electron micrograph of our $p$-GaAs nanowire MESFET structure. It features a $\sim 120$~nm diameter GaAs nanowire with an undoped core (light green in {\bf a}) and a nominally $30$~nm thick Be-doped shell (dark green). The nominal shell acceptor density $N_{A} = 1.5 \times 10^{19}$~cm$^{-3}$. The nanowire is thinned near the middle of its length via a lithographically patterned wet-etch to facilitate Schottky-gating as described in the text. The device has three electrodes (yellow): source (S), drain (D) and a Schottky-gate (G). The typical gate length for our devices is $600-750$~nm. The scale bar in {\bf b} represents $1~\mu$m.}
\end{figure}

Here we report on the development of a $p$-GaAs nanowire metal-semiconductor field-effect transistor (MESFET) with sub-threshold swing $S = 62$~mV/dec, normalised peak transconductance $g_m~=~1.2~\mu$S$/\mu$m, on-off ratio $10^{5.1}$, on-resistance as low as $715$~k$\Omega$, contact resistance $\sim30$~k$\Omega$, a field-effect hole mobility $\sim 5$~cm$^2$/Vs and high-fidelity operation at frequencies up to $10$~kHz. Two notable aspects of GaAs nanowires are the capacity for direct self-catalysed growth,~\cite{FontcubertaAPL08} including on Si substrates,~\cite{KrogstrupNL10, CasadeiAPL13} and surface-states that pin the surface Fermi energy to mid-gap giving metal-semiconductor interfaces with a substantial Schottky barrier, typically $\sim~0.5$~eV.~\cite{WaldropAPL84} The latter can be both an advantage and a disadvantage. On the positive side, it enables us to make Schottky gates. This simplifies the device processing as the gates become self-insulating, removing the need for an added insulator layer, e.g., Al$_2$O$_3$ or HfO$_2$. The disadvantage is the difficulty involved in achieving low-resistance ohmic contacts to $p$-GaAs nanowires. Low resistance contacts ($\sim 30$~k$\Omega$) can be achieved using GaAs nanowires with a heavily Be-doped shell,~\cite{UllahNanotech17} however, such heavy doping means conventional metal-oxide gating fails.~\cite{UllahPRM18} Here we demonstrate that this problem can be overcome by carefully etching the nanowire to reduce the local Be-doping density at the locations where gates are patterned prior to depositing gate metal directly on the nanowire surface. Our approach provides both strong, low-leakage gating and low contact resistance simultaneously. It opens a path to high performance $p$-GaAs nanowire transistors without oxide gate-insulators, which entail substantial issues with charge trapping and gate hysteresis.~\cite{DayehAPL07, RoddaroAPL08, HollowayJAP13} Our $p$-GaAs nanowire MESFETs could potentially be paired with $n$-InGaAs~\cite{NoborisakaJJAP07}, $n$-GaAs~\cite{FortunaEDL09} or $n$-AlGaAs/GaAs MESFETs~\cite{MorkotterNL15} to produce high-performance oxide-free complementary circuit architectures.

\begin{figure}
\includegraphics[width=12cm]{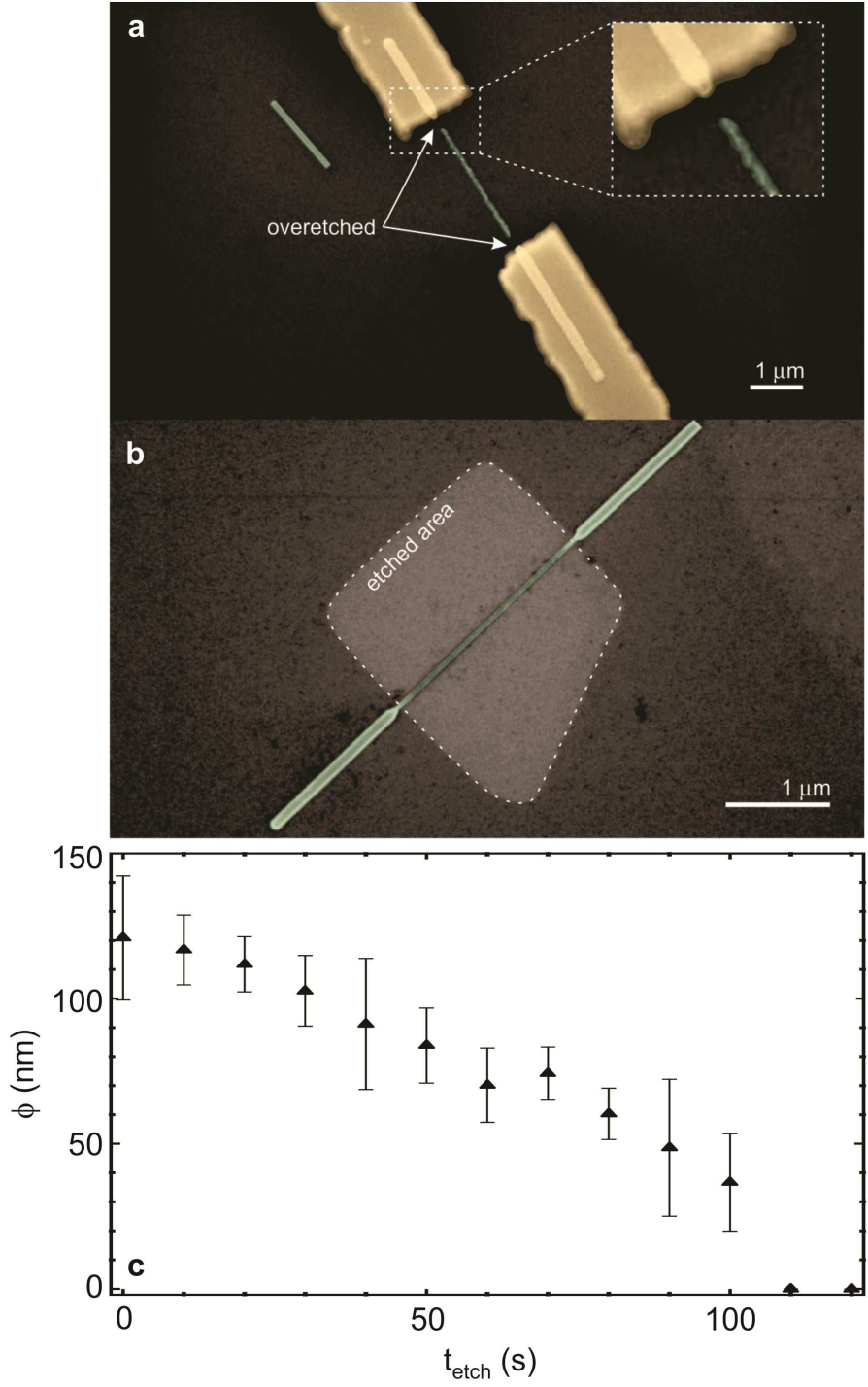}
\vspace{2mm}
\caption{{\bf Controlled etching of the Be-doped nanowire shell.} Scanning electron micrographs of a GaAs nanowire etched with $1:1:250$ H$_3$PO$_4$:H$_2$O$_2$:H$_2$O for {\bf a} self-aligned etch using the source/drain contacts as etch mask and {\bf b} patterned etch with PMMA resist etch mask demonstrating issues with metal-assisted etching for this process. Both scale bars represent $1~\mu$m. {\bf c} Plot of etched region diameter $\phi$ vs etch time $t_{etch}$ for the $1:1:250$ H$_3$PO$_4$:H$_2$O$_2$:H$_2$O etchant solution on a GaAs nanowire without source/drain contacts.}
\end{figure}

{\bf A controlled near-monolayer accuracy etch for GaAs nanowires} Figure~1 shows a schematic of the $p$-GaAs nanowire MESFET structure and a scanning electron micrograph of a completed device. The nanowire is grown as an undoped core with a heavily Be-doped shell (see Methods for full details), however, this does not mean the Be-dopants remain confined to the shell alone. Be has a high diffusion constant in GaAs,~\cite{IlegemsJAP77} which means significant amounts of the Be diffuse into the otherwise-undoped core.~\cite{CasadeiAPL13} The result is a roughly constant doping density in the shell and a monotonically decreasing doping density moving into the core;~\cite{CasadeiAPL13} for a much higher doping density than we use, this can result in a fully doped core.~\cite{KorenNL11} This aspect of the doping profile is crucial to properly understanding our device. It also underpins our ability to make devices with high repeatability and yield given a well-engineered nanowire etch process. We leave the shell intact at the source and drain contacts as shown in Fig.~1 to minimise the contact resistance. The key is then to thin the nanowire and thereby reduce the doping level at the gate location just enough for the channel to become electrostatically gateable in the region where the gates are placed. This requires a careful balance -- etch insufficiently and the gate either fails to function~\cite{UllahPRM18} and/or leaks current to the nanowire; etch too much and the on-resistance become undesirably high and the on-off ratio falls.~\cite{UllahNanotech17} Note that it is vital that some doping remains in the etched region to counteract the GaAs surface-states, which pin the Fermi energy at mid-gap, and then provide the free holes required to achieve conduction through the etched channel segment. The gate-etch is thus administered completely ignoring the location of the core-shell boundary. Indeed, for our optimum devices below, we completely etch the shell and somewhat into the core.

We used a phosphoric acid etch $1:1:250$ H$_3$PO$_4$:H$_2$O$_2$:H$_2$O, inspired by Mori and Watanabe's work~\cite{MoriJECS78} on slow etching of GaAs wafers for microwave devices to obtain a highly-controllable nanowire gate-etch without deleterious effects. We made attempts with other etchants, e.g., sulfur-oleylamine, which has previously shown near-monolayer etch control for III-Vs~\cite{NaureenAFM13}, but only the phosphoric acid etchant proved compatible with other aspects of nanowire device fabrication (see Supplementary Information for further discussion). We note that a two-step digital etch process could also be applied;~\cite{LuEDL17} it may provide more precision than our \ce{H3PO4} etch but our $30-50$~nm etch depth may also make it a high time-cost option. With this in mind, one could suggest using a thinner shell and increasing the doping density to avoid loss in contact performance. The issue here is that this drives dopant diffusion into the core,~\cite{KorenNL11} as discussed above, requiring a deeper gate-etch to achieve a functional doping level at the gated segment. This defeats the purpose of thinning the shell in the first place.

An important consideration when performing the gate-etch for a GaAs nanowire MESFET is the stage where the etch is performed. The simplest option is to pattern the source and drain contacts first and use these as a mask for a `self-aligned' gate-etch~\cite{KaneAPL93}. The outcome is shown in Fig.~2a, where the nanowire etches much faster local to the contact edges than near the middle, presumably due to metal-assisted etching~\cite{DeJarldNL11}. This is supported by the fact that the nanowire segment off to the left in Fig.~2a, which was also exposed to the etchant, is not as heavily etched. We avoided this problem by performing the etch prior to contact deposition, using EBL-patterned polymethylmethacrylate (PMMA) resist as the etch mask. This gives the more even gate-etch shown in Fig.~2b, and was the etch process used to obtain the device shown in Fig.~1b. The fact that the nanowire is sitting on a substrate may slightly reduce the etch rate at the bottom of the nanowire, which could produce a slight inhomogeneity in doping density on this side. We do not expect this effect to be substantial given the same issue did not adversely affect our earlier horizontal wrap-gate nanowire devices,~\cite{StormNL12, BurkeNL15} where a similar wet-etch was administered, or appear to be detrimental to the performance of the devices made here. This issue would be eliminated entirely in the vertical orientation preferred for applications.

\begin{figure*}
\includegraphics[width=17cm]{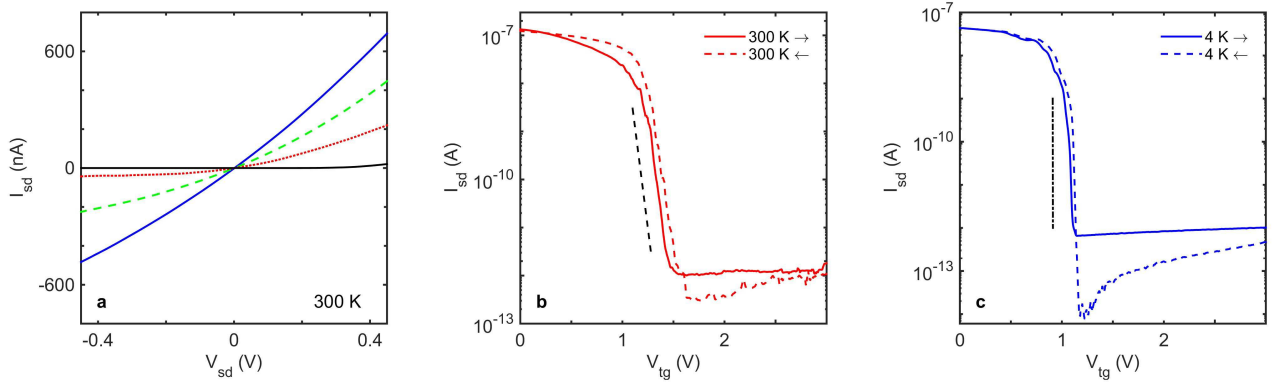}
\vspace*{2mm}
\caption{{\bf $p$-GaAs nanowire MESFET dc performance.} {\bf a} Source-drain current $I_{sd}$ vs source-drain bias $V_{sd}$ at four different top-gate voltages $V_{tg} = +0$~V (solid blue), $+0.5$~V (dashed green), $+1.0$~V (dotted red) and $+1.5$~V (thick black) for a $p$-GaAs MESFET with $40$~s gate-etch at temperature $T = 300$~K. {\bf b/c} $I_{sd}$ vs top-gate voltage $V_{tg}$ at {\bf b} $T = 300$~K and {\bf c} $T = 4$~K with sweep to positive $V_{tg}$ (solid) and return to $V_{tg} = 0$~V (dashed) both shown. The thermal limit sub-threshold swings at $T = 300$~K ($59.6$~mV/dec) and $T = 4$~K ($0.8$~mV/dec) are shown as the dashed black line in {\bf b} and the dot-dashed black line in {\bf c} for reference. The gate-voltage separation between up- and down-sweeps at the sub-threshold midpoint $I_{sd} = 10^{-10}$~A is $80$~mV for {\bf b} and $40$~mV for {\bf c}. The apparent hysteresis in the off-state for {\bf b} and {\bf c} is an experimental artefact explained in the Supplementary Information.}
\end{figure*}

Figure~2c shows a plot of the etched segment diameter $\phi$ versus etch time $t_{etch}$ for the phosphoric acid etch. The diameter was measured by post-etch scanning electron microscopy, with results averaged over $24$ separate nanowires for each $t_{etch}$. The etch is well controlled for $t_{etch} < 100$~s. At longer $t_{etch}$ the nanowire becomes sufficiently thin that the etch acts anisotropically, producing breaks and substantial thickness variations. The long $t_{etch}$ behaviour is not problematic since we only aim to slightly thin the nanowire and this requires significantly less than $t_{etch} = 100$~s. Some readers may note a slight gradient change to the linear trend in Fig.~2c at approximately $30$~s. Given a nominal shell thickness of $30$~nm and a typical etch rate of $1.1$~nm/s, this likely arises from the etch crossing the core-shell boundary.

\begin{figure}
\includegraphics[width=12cm]{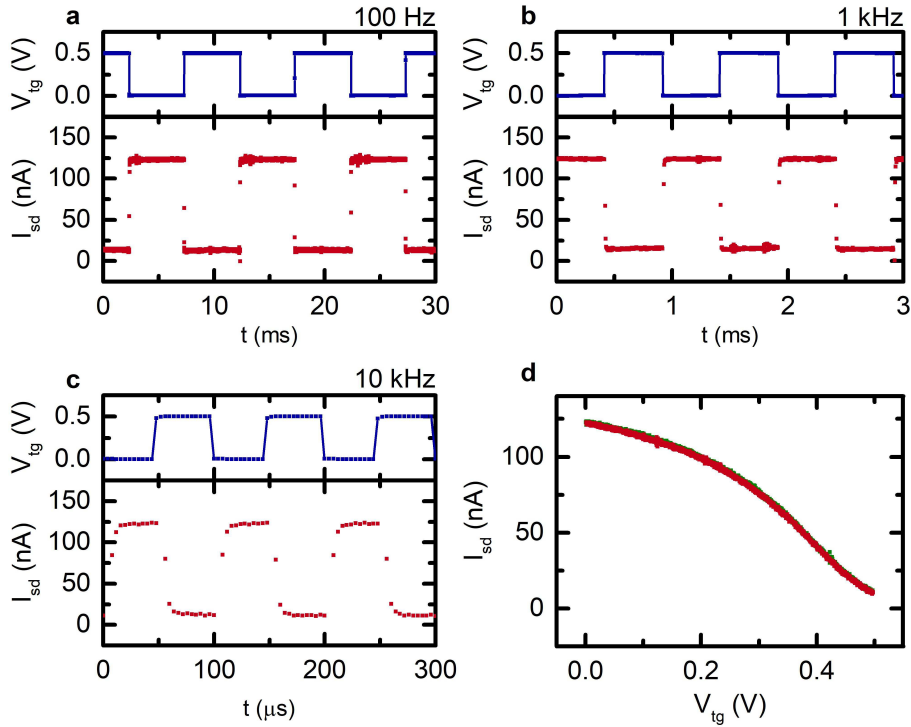}
\vspace{2mm}
\caption{{\bf p-GaAs nanowire MESFET ac performance.} {\bf a-c} Plots of applied gate voltage $V_{g}$ (blue/top panel) and source-drain current $I_{sd}$ (red/bottom panel) vs time $t$ for gate drive frequencies $f$ of {\bf a} $100$~Hz, {\bf b} $1$~kHz and {\bf c} $10$~kHz. {\bf d} source-drain current $I_{sd}$ vs gate voltage $V_{g}$ for increasing $V_{g}$ (red) and decreasing $V_{g}$ (green) demonstrating the minimal gate hysteresis in this device for $f = 100$~Hz operation (n.b. hysteresis is sufficiently small that the green trace is hidden by the red trace in Fig.~5{\bf d}).}
\end{figure}

{\bf Obtaining high-performance $p$-GaAs nanowire MESFETs with controlled gate-segment etching} The challenge in the fabrication is to identify the appropriate $t_{etch}$ needed to optimise electrical performance. Optimum performance was obtained for $t_{etch} = 40$~s, yielding the electrical characteristics in Fig.~3. Figure~3a shows $I_{sd}$ versus $V_{sd}$ at four different top-gate voltages $V_{tg}$. The $I_{sd}$ versus $V_{sd}$ characteristic is relatively linear for $V_{tg} = 0$~V, with slope corresponding to an unbiased channel resistance $R_{on}\sim700$~k$\Omega$. The typical contact resistance at $T = 300$~K for these nanowires is $29 \pm 15$~k$\Omega$~\cite{UllahNanotech17}. An increasingly positive $V_{tg}$ results in reduced $I_{sd}$ at a given $V_{sd}$, as expected for a $p$-MESFET, without severe degradation in source-drain characteristics. Figure~3b shows $I_{sd}$ versus $V_{tg}$ at $V_{sd} = 100$~mV at $T = 300$~K. We obtain remarkably strong gating with a sub-threshold swing of $62 \pm 7$~mV/dec at $T = 300$~K. This is within $4\%$ of the room-temperature thermal limit of $59.6$~mV/dec, which is indicated by the black dashed line in Fig.~3b for comparison. The $T = 300$~K data has on-off ratio $10^{5.1}$, threshold voltage $V_{th} = +1.3$~V and relatively low hysteresis; the up and down traces in Fig.~3b are separated by only $80$~mV in $V_{tg}$ at $I_{sd} = 10^{-10}$~A. We will comment further on normalised on-current, peak transconductance and field-effect mobility in our benchmarking comparison to other devices below.

{\bf Potential for studying hole-based quantum devices} Our device architecture may also be useful for quantum device applications. A concern is that gate-etch might cause conduction to freeze-out before sufficiently low temperature for observing quantum effects is attained. To test this we repeated the $I_{sd}$ versus $V_{tg}$ characterisation after cooling to $T = 4$~K, as shown in Fig.~3c. The on-resistance increases slightly to $R_{on}~\sim~1.6$~M$\Omega$ consistent with the etched channel segment being non-metallic. Strong gating is retained but the sub-threshold swing of $15 \pm 7$~mV/dec is not as close to the respective thermal limit ($0.8$~mV/dec at $T = 4$~K -- black dot-dashed line in Fig.~3c for comparison). This is a positive outcome for quantum device applications, as very steep gate response makes it harder to control/study quantum effects, e.g., 1D conductance quantization~\cite{vanWeperenNL13, HeedtNL16, IrberNL17} and can increase noise as smaller gate voltage fluctuations give a stronger effect. The on-off ratio at $T = 4$~K remains high ($10^{4.6}$) and the threshold voltage $V_{th} = +1.1$~V shifts slightly closer to zero, consistent with reduced ionized dopant density due to freeze-out. The hysteresis is also improved significantly upon cooling. Optimisation of the gate-etch for quantum device applications will be subject of a separate study.

{\bf Comparison of performance with $p$-GaSb nanowire MOSFETs} The most notable competitor for our devices is the $p$-GaSb nanowire MOSFET, which is the current device of choice for complementary architectures featuring III-V nanowires. We will make this comparison in two stages. First we will compare to horizontally-oriented single nanowire devices to make a direct `like-for-like' performance comparison. We will then benchmark against more recently developed vertical nanowire array devices to consider prospects for scale-up and applications.

Dey {\it et al.}~\cite{DeyNL12} previously reported single InAs/GaSb nanowire CMOS inverter oriented horizontally on a \ce{SiO2}-on-Si substrate. Their $p$-GaSb MOSFET featured a nanowire with diameter $65$~nm and a gate length of $\sim 500$~nm, giving sub-threshold swing $S = 400$~mV/dec, peak transconductance $g_{m} = 3.4~\mu$S$/\mu$m, on-off ratio $\sim 10^{1.8}$ and a normalised on-resistance $R_{on,n} = 212$~$\Omega$.mm at room temperature. Babadi {\it et al.}~\cite{BabadiAPL17} improved $R_{on,n}$ to $3.9$~$\Omega$.mm by moderate doping with Zn coupled with a reduced gate length of $200$~nm. This improved the peak transconductance to $g_m = 80~\mu$S$/\mu$m but also led to a significantly reduced sub-threshold swing of $820$~mV/dec.~\cite{BabadiAPL17} A measurement of hole mobility was not provided in Dey {\it et al.}~\cite{DeyNL12} but Babadi {\it et al.}~\cite{BabadiAPL17} obtained a peak hole mobility of $153$~cm$^2$/Vs.

Our sub-threshold swing of $62$~mV/dec significantly surpasses that obtained by Dey {\it et al.} and Babadi {\it et al.} above. Our peak transconductance $g_m = 1.3~\mu$S$/\mu$m (normalised to channel circumference) is comparable with that obtained by Dey {\it et al.}~\cite{DeyNL12} Our normalised on-resistance is $137$~$\Omega$.mm, which is $65\%$ lower than the device reported by Dey {\it et al.} but $\sim 35$ times higher than Babadi {\it et al.}~\cite{BabadiAPL17} Improvement in our normalised on-resistance should be possible by reducing the channel length and/or increasing the shell doping density to improve the contact resistance and then compensating with increased gate-etch depth to maintain gate-performance. The hole mobility is challenging to calculate for our device due to the absence of an oxide, but we can provide a reasonable estimate. The gate insulator in our device is the depletion region at the GaAs surface arising from surface states. Casadei {\it et al.}~\cite{CasadeiAPL13} calculate the surface depletion width as a function of doping density for nanowire radii of $40$~nm and $100$~nm. They obtain a depletion width of $7$~nm for our acceptor density $N_{A} = 1.5 \times 10^{19}$~cm$^{-3}$ with negligible radii dependence, i.e., the depletion width values as a function of radii converge at high $N_{A}$.~\cite{CasadeiAPL13} To obtain the top-gate capacitance we use a coaxial-capacitor model and reduce the resulting capacitance by $20\%$ to account for the gate not covering the bottom of the nanowire where it sits on the substrate. Assuming the standard value for GaAs dielectric constant of $12.9$ and the measured gate-length of $650$~nm for the device with $t_{etch} = 40$~s, we obtain a gate capacitance of $1.8$~fF. This in turn gives a hole mobility of $5.6$~cm$^2$/Vs. This is lower than the mobility obtained by Babadi {\it et al.}~\cite{BabadiAPL17} from devices with less than a third of the gate-length and much lower on-resistance. More competitive mobilities are probably attainable for $p$-GaAs MESFETs with some optimisation of our design.

Figure~4 shows the frequency response of our $t_{etch} = 40$~s $p$-GaAs MESFET. A square-wave top-gate voltage $V_{tg}$ (blue trace/top panels) is applied at frequency $f = 100$~Hz, $1$~kHz and $10$~kHz producing a square-wave source-drain current $I_{sd}$ response (red trace/bottom panels) in each case. The fidelity for our device remains excellent up to $10$~kHz, where the leading edge begins to evolve a slight rounding. The high-frequency fidelity loss does not arise from the device acting as a low-pass filter -- the $1.8$~fF gate capacitance estimated above combined with the $715$~k$\Omega$ channel resistance gives a roll-off frequency of $123$~MHz. That said, radio-frequency operation is well-known to require careful capacitance management of the device and external circuit, and is commonly achieved for nanowire transistors by moving to vertical nanowire array structures.~\cite{WernerssonProcIEEE10, JohanssonEDL14} The fidelity loss we see likely arises from a mixture of instrument/circuit limitations and some gate-semiconductor interface effects; there will inevitably be some GaAs native oxide at the gate interface because the processing is not carried out in a fully oxygen-free environment. Charge trapping by oxide at gate interfaces is a well-known contributor to gate hysteresis in nanowire transistors.~\cite{DayehAPL07, RoddaroAPL08, HollowayJAP13} The data in Fig.~3b indicates low hysteresis under dc operation; this performance is retained under ac conditions also. Figure~4d shows the device response to a triangle-wave $V_{tg}$ signal between $0.0$ and $+0.5$~V at $f = 100$~Hz with the up-sweep to $+0.5$~V data in green and the down-sweep data in red. Up and down sweeps are separated by a constant offset at higher frequencies (see Supplementary Fig.~S3). We attribute this to a displacement current induced by the high gate-sweep rate. Regarding benchmarking of the frequency performance against $p$-GaSb, Dey {\it et al.}~\cite{DeyNL12} only present characteristics for their inverter and not their $p$-GaSb MOSFET alone unfortunately. However, their inverter frequency response will be limited by the slowest component, which is inevitably the $p$-GaSb MOSFET, giving some indication of its performance. The frequency performance of our $p$-GaAs MESFET appears at least as good by this comparison -- a more conclusive benchmarking would require a careful comparative study beyond the scope of the current work.

A notable aspect of our devices is the low gate leakage current despite the lack of an oxide gate insulator. We obtain gate leakage current $I_{tg}~<~50$~pA over the entire operating range for devices with $t_{etch} = 40$~s. The gate-etch is essential to obtaining low gate leakage; for $t_{etch} = 0$~s we get $I_{tg} \sim 10$~nA at $V_{tg} = +2$~V (see Supplementary Fig.~S4), as expected, since in this instance the heavy shell doping makes the metal gate electrode into an ohmic contact~\cite{BacaTSF97}. For $t_{etch} > 40$~s we get $100\%$ yield of leakage-free gates albeit with reduced gate performance -- Data obtained for $t_{etch} = 50$~s and $60$~s corresponding to Figs.~3 and 4 is presented in Supplementary Figs.~$S5-8$ to demonstrate this. At $t_{etch} = 40$~s we get excellent performance but the yield of leakage-free gates is only $\sim70\%$ and drops sharply for shorter etch times. This makes $t_{etch} = 40$~s optimum on both performance and yield-to-performance measures, however, this would naturally vary for different growth conditions and require optimisation on a growth-batch basis. This performance compares very well to other single nanowire MESFETs, for example, Noborisaka {\it et al.}~\cite{NoborisakaJJAP07} report leakage currents between $1$ and $700$~pA for a $n$-InGaAs device, Fortuna \& Li~\cite{FortunaEDL09} report $20$~pA to $30$~nA for a $n$-GaAs device, and Liu {\it et al.}~\cite{LiuJAP08} report $10-20$~pA for a $p$-\ce{Zn3P2} device.

{\bf Prospects for scale-up} Development for practical applications tends to focus on vertical nanowire arrays over horizontal single nanowire devices for reasons of higher current carrying capacity, improved gate coupling and scalability, and capacitance management for high-frequency operation.~\cite{WernerssonProcIEEE10, RielMRSBull14} A key aspect in vertical array devices is that the gate-length is no longer constrained by lithography and instead controlled by layer thicknesses during processing, facilitating scaling to sub-$100$~nm gate length. This pathway has already been taken for $p$-GaSb, with the single horizontal nanowire devices of Dey {\it et al.}~\cite{DeyNL12} and Babadi {\it et al.}~\cite{BabadiAPL17} translated into vertical nanowire array structures by, e.g., Svensson {\it et al.}~\cite{SvenssonNL15} and J\"{o}nsson {\it et al.}~\cite{JonssonEDL18}. The nanowire III-V CMOS on Si inverter structures by J\"{o}nsson {\it et al.}~\cite{JonssonEDL18} represent the state-of-the-art in the field and feature a $p$-GaSb device containing an array of $144$ nanowires with $350$~nm pitch, $40$~nm diameter and $70$~nm gate length. These transistors give sub-threshold swing $273$~mV/dec, normalised on-resistance $5.9~\Omega$.mm and peak transconductance $74~\mu$S$/\mu$m. The on-resistance and peak transconductance are superior, as one would expect for an array. Our sub-threshold swing is better, but we have to acknowledge that this metric is often compromised slightly in array structures by wire-to-wire variations. Benchmarking on ac operation is difficult because such data is limited for $p$-GaSb array transistors. Svensson {\it et al.}~\cite{SvenssonNL15} demonstrate ac operation of an inverter featuring $n$-InAs and $p$-GaSb vertical nanowire array transistors at $1$~kHz with good fidelity but report significant distortion at higher frequencies due to parasitic capacitance issues. These issues are more related to the vertical nanowire array layout than the nanowire channel itself; at this point the only comment we can make is that our $p$-GaAs nanowire performance does not indicate that scale-up to arrays for high frequency operation would be obviously problematic.

{\bf Potential for complementary GaAs nanowire MESFET circuits} Looking forwards, an interesting consideration is the scope for complementary circuits featuring only (In,Al)GaAs nanowire MESFETs, which would require an $n$-channel device to complement the $p$-GaAs design we present here. We see several alternatives. The first is a pure GaAs implementation. Fortuna and Li~\cite{FortunaEDL09} reported a $n$-GaAs nanowire MESFET featuring a nanowire grown epitaxially along the surface of a $(100)$GaAs substrate. Their device gave strong electronic performance with sub-threshold swing $S = 150$~mV/dec, on-off ratio $\sim 10^{2}$ and low contact resistance $R_{c} \sim 40$~k$\Omega$. Noborisaka {\it et al.}~\cite{NoborisakaJJAP07} reported an $n$-InGaAs nanowire MESFET with sub-threshold swing $S = 200$~mV/dec, on-off ratio $\sim 10^{3}$ and low channel resistance $R_{on} \sim 100$~k$\Omega$. More recently, Mork\"{o}tter {\it et al.}~\cite{MorkotterNL15} demonstrated a delta-doped GaAs-AlGaAs core-shell nanowire $n$-MESFET with $S = 70$~mV/dec, on-off ratio exceeding $10^{4}$ and $R_{c} \sim 30$~k$\Omega$. With the latter, complementary circuits could be made on Si substrates using `pick and place' micro-manipulation techniques. An exciting alternative would be to grow the GaAs nanowires directly on Si by template-assisted selective epitaxy~\cite{KnoedlerCGD17}, although a challenge here would be doping of separate GaAs nanowires to be deterministically $n-$ or $p$-type.

We demonstrated a $p$-GaAs nanowire MESFET with strong electronic performance by taking a GaAs nanowire with an undoped core and heavily Be-doped shell and carefully etching the nanowire at the gate location to enable a Schottky-gate without compromising on contact resistance. We obtain a sub-threshold swing of $62 \pm 7$~mV/dec, within $4\%$ of the room-temperature thermal limit, on-off ratio $\sim 10^{5}$, on-resistance $\sim 700$~k$\Omega$ and typical contact resistance $\sim 30$~k$\Omega$. Accounting for nanowire diameter, we obtain a normalised on-resistance of $137$~$\Omega$.mm and peak transconductance of $g_m = 1.3~\mu$S$/\mu$m. We estimate a field-effect hole mobility of $\sim 5$~cm$^2$/Vs for our device. Our overall single MESFET performance is competitive with that obtained for single horizontal $p$-GaSb nanowire MOSFETs,~\cite{DeyNL12,BabadiAPL17} and suggests that scale-up to vertical nanowire transistor arrays may lead to devices that could potentially compete well with $p$-GaSb vertical nanowire array transistors.~\cite{SvenssonNL15, JonssonEDL18} Our $p$-GaAs MESFETs show good square-wave fidelity for frequencies up to $10$~kHz, comparable to single horizontal $p$-GaSb nanowire MOSFETs~\cite{DeyNL12}. The ability to eliminate the gate oxide in our device results in strong gating and reduced issues with charge-trapping effects, e.g., gate hysteresis. Our $p$-GaAs MESFETs show strong potential for combination with high-performance $n$-type (In,Al)GaAs MESFETs~\cite{NoborisakaJJAP07, FortunaEDL09, MorkotterNL15} towards nanowire complementary circuit applications, perhaps even ultimately integrated monolithically on Si using templated epitaxy techniques~\cite{KnoedlerCGD17}.

{\bf Methods}\\
The GaAs nanowires were self-catalysed~\cite{ColomboPRB08} and grown by molecular beam epitaxy on $(111)$Si~\cite{CasadeiAPL13}. The undoped core was grown at $630^{\circ}$C using As$_4$ and a V/III flux ratio of $60$ for $30 - 45$~min. The Be-doped shell was grown at $465^{\circ}$C using As$_2$ and a V/III ratio of $150$ for $30$~min. giving nanowires with shell acceptor density $N_{A} = 1.5 \times 10^{19}$~cm$^{-3}$, diameter $120 \pm 20$~nm and length $5-7~\mu$m. These nanowires are the highest doping density from earlier work on contacts to $p$-GaAs nanowires~\cite{UllahNanotech17}. The nanowires are pure zincblende crystal phase but may have short wurtzite segments at the ends~\cite{KrogstrupNL10}; these wurtzite segments will be buried under the source/drain contacts. Nanowires were dry-transferred to a pre-patterned HfO$_2$/SiO$_2$-coated $n^+$-Si substrate for device fabrication. The heavily-doped substrate was used as a global back-gate for control studies. The source and drain contacts were defined by electron-beam lithography (EBL) and thermal evaporation of $200$~nm of $1:99$ Be:Au alloy (ACI alloys) and were not thermally annealed~\cite{UllahNanotech17}. The thin GaAs native oxide was removed at the contact interfaces by a $30$~s etch in $10\%$ HCl solution immediately prior to contact deposition. The $20$/$180$~nm Ti/Au gate electrode was formed in a separate round of EBL and metal deposition. Electrical measurements were performed at temperature $T = 300$~K in ambient atmosphere and at $T = 4$~K by immersion in liquid helium. The source-drain current $I_{sd}$ was measured using a Keithley 6517A electrometer for the dc measurements in Fig.~3 and using a Femto DLPCA-200 pre-amplifier connected to a National Instruments USB-6216 data acquisition system for the ac measurements in Fig.~4. Keithley K2410 voltage sources supplied the source-drain voltage $V_{sd}$, top-gate voltage $V_{tg}$ and back-gate voltage $V_{bg}$ (back-gate used in control studies only) for the dc measurements. In the ac measurements $V_{tg}$ was instead supplied from a Stanford Research Systems DS345 signal generator.

{\bf Supporting Information.} Additional fabrication details and electrical data, as well as a discussion of attempts with sulfur-oleylamine etchant. This material is available free of charge via the Internet at http://pubs.acs.org.

\acknowledgement

We thank D.J. Carrad for helpful discussions. This work was funded by the Australian Research Council (ARC) grants DP170102552 and DP170104024, the University of New South Wales, Danish National Research Foundation and the Innovation Fund Denmark. This work was performed in part using the NSW node of the Australian National Fabrication Facility (ANFF).

\end{document}